# An Oscillation Evident in Both Solar Neutrino Data and Radon Decay Data

P.A. Sturrock[1], E. Fischbach[2], O. Piatibratova[3], G. Steinitz[3], and F. Scholkmann[4]

P.A. Sturrock
    sturrock@stanford.edu
E. Fischbach
    Ephraim@physics.purdue.edu
O. Piatibratova
    oksana@gsi.gov.il
G. Steinitz
    steinitz@gsi.gov.il
F. Scholkmann
    felix.scholkmann@googlemail.com

[1] Center for Space Science and Astrophysics and Kavli Institute for Particle Astrophysics and Cosmology, Stanford University, Stanford, CA 94305-4060, USA;
[2] Department of Physics and Astronomy, Purdue University, West Lafayette, IN 47907, USA;
[3] Geological Survey of Israel, Jerusalem, 95501, Israel;
[4] Research Office for Complex Physical and Biological Stems, Zurich, Switzerland.

We regret to inform our readers that Gideon Steinitz passed away in the spring of 2018.

**Abstract**

Analyses of neutrino measurements acquired by the Super-Kamiokande Neutrino Observatory (SK, for the time interval 1996 - 2001) and of radon decay measurements acquired by the Geological Survey of Israel (GSI, for the time interval 2007 - 2017) yield remarkably consistent detections of the same oscillation:

| | | |
|---|---|---|
| frequency | 9.43 ± 0.04 year$^{-1}$ (SK), | 9.44 ± 0.04 year$^{-1}$ (GSI); |
| amplitude | 6.8 ± 1.7 % (SK), | 7.0 ± 1.0 % (GSI); |
| phase | 124 ± 15 deg. (SK), | 124 ± 9 deg. (GSI). |

We briefly discuss possible hypotheses that may be relevant to this experimental result.

## 1. Introduction

This article is concerned with an apparent interplay between the beta-decay process and neutrinos. Evidence for the variability of the beta-decay process has been presented in earlier articles [1-3]. We now address the suggestion that the beta-decay process may be influenced by solar neutrinos [4-7]. We here present evidence that an oscillation of frequency 9.43 year$^{-1}$, previously recognized in Super-Kamiokande solar neutrino data [8, 9], is recognizable also in certain nuclear decay measurements. We focus once more on radon decay measurements acquired at the Geological Survey of Israel GSI), which comprise the most extensive compilation of nuclear decay measurements [10,11]. However, we note that this oscillation has also been detected by Alexeyev et al. in their measurements of the decay of $^{213}$Po and of $^{214}$Po [12].

## 2. Super-Kamiokande Measurements

Figure 1 shows the power spectrum derived from Super-Kamiokande flux measurements for the interval 1996.4 to 2001.6 [8].

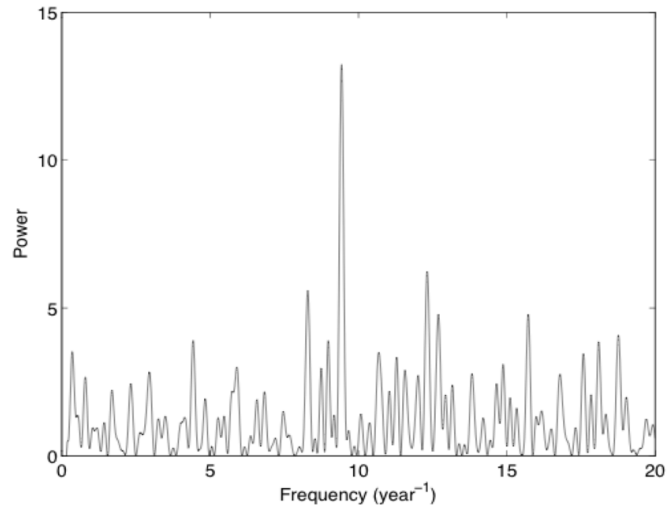

Figure 1. The power spectrum formed from Super-Kamiokande data, taking account of start time, end time, flux estimate, lower error estimate and upper error estimate.

This power spectrum was derived from Super-Kamiokande data [13] by a likelihood procedure that takes account of the start time, end time, flux estimate, lower error estimate and upper error estimate of each time bin. (By comparison, the original Super-Kamiokande analysis [13] took account only of the mean time and mean flux estimate of each bin.) According to this analysis [8], the power spectrum, shown in Figure 1, has its maximum value, S = 13.24, at 9.43 year$^{-1}$. (This oscillation has also been identified by Ranucci [9], who made a slightly different selection of information that led to an oscillation with power 10.84 at 9.42 year$^{-1}$.)

The standard formula for the probability $P$ of obtaining a peak of power $S$ or more from (assumed normally distributed) random fluctuations [14] is

$$P = e^{-S}. \qquad (1)$$

The accuracy of this estimate for the current application has been verified by Monte Carlo simulation [8]. We see that the probability of finding a peak of power 13.24 or more at a specified frequency is $2 \cdot 10^{-6}$.

We have also determined the probability of finding by chance a peak of this power or more anywhere in a search band appropriate for internal solar rotation, which - based on helioseismology [15] - we take to be 6 – 16 year$^{-1}$. This value was found to be $10^{-4}$.

### 3. The GSI Radon Experiment

The Geological Survey of Israel (GSI) experiment [1,2,3,10,11] has recorded measurements of gamma photons and alpha particles arising from radon decay every 15 minutes from day 86 of 2007 to day 312 of 2016. In the present article, we analyze only gamma-ray measurements packaged into 1-hour bins, comprising approximately 88,000 lines of data. It is relevant to note that, as we see from Figure 2 of ref. [7], the gamma detector is located vertically above the region in which radon decay occurs. The mean count per 1-hour bin is $9.42 \cdot 10^5$.

Figures 2 and 3 show the power spectra obtained from measurements acquired for the frequency bands 0 – 6 year$^{-1}$ and 6 – 16 year$^{-1}$, respectively. In each figure we show the power spectra derived from noontime data (10 am to 2 pm, local time) and from midnight data (10 pm to 2 am, local time).

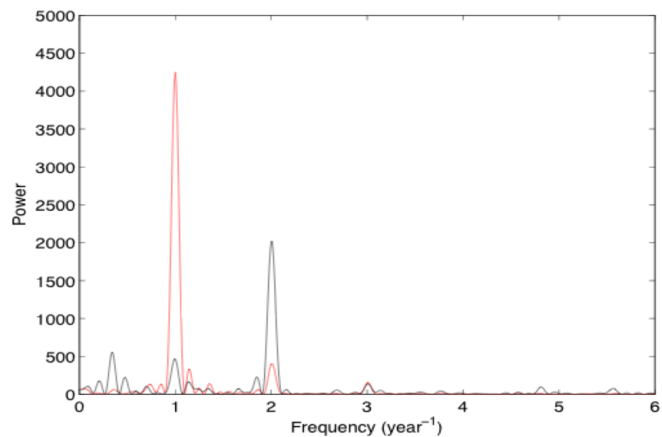

Figure 2. Power spectra formed from the 4-hour bands of measurements centered on noon (red) and midnight (blue) for the frequency band 0 – 6 year$^{-1}$. We see that the strongest daytime oscillation is at 1 year$^{-1}$; the strongest nighttime oscillation is at 2 year$^{-1}$.

We see from Figure 2 that, for each time interval, there are exceedingly strong annual and biennial oscillations. We see from Figure 3 that oscillations in the solar-rotation search band 6 – 16 year$^{-1}$ are clearly evident in the midnight data, but less evident in the noon data. This distinction is compatible with the conjecture that the influence of neutrinos on nuclear decay is such that emergent gammas tend to travel in the same direction as the incoming neutrinos [7].

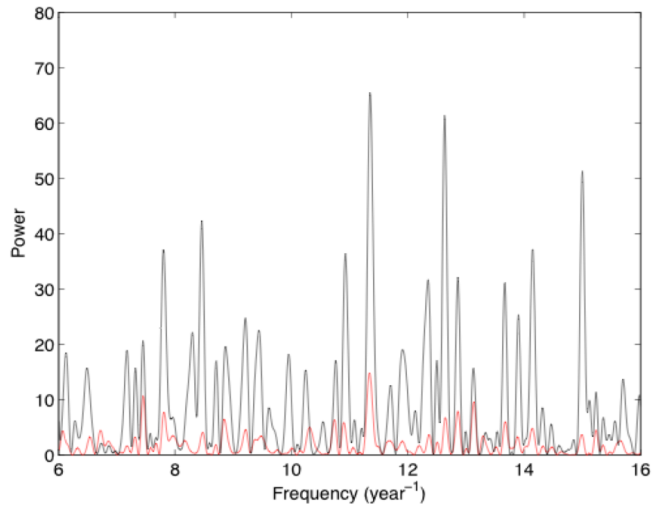

Figure 3. Power spectra formed from the 4-hour bands of measurements centered on noon (red) and on midnight (blue) for the frequency band 6 – 16 year$^{-1}$. We see that there are strong oscillations in the frequency band 11 – 13 year$^{-1}$ (note especially the peaks at 11.35 year$^{-1}$ and 12.64 year$^{-1}$) in the nighttime data, but comparatively small oscillations in the daytime data.

A list of the top 20 peaks in the power spectrum formed from the midnight data, for the frequency band 6 – 16 year$^{-1}$, is shown in Table 1. We see that there is a peak at 9.44 year$^{-1}$ with power S = 22.6. Since the mean separation of the peaks is approximately 0.17 year$^{-1}$, this frequency is indistinguishable from that (9.43 year$^{-1}$) in the Super-Kamiokande power spectrum. According to Equation (1), the probability that the peak at 9.44 year$^{-1}$ with power 22.6 is attributable to normally distributed random fluctuations is 1.5 10$^{-10}$.

For later reference, we draw attention to two annual sidebands (companion peaks offset by multiples of 1 year$^{-1}$) at 7.45 year$^{-1}$ (with power 20.7) and at 8.46 year$^{-1}$ (with power 42.4). Such sidebands are known to be characteristic of oblique rotators [16].

Table 1. Top 20 peaks in the power spectrum formed from midnight data in the frequency band 6 – 16 year$^{-1}$.

| Frequency (year$^{-1}$) | Power | Order |
|---|---|---|
| 6.13 | 18.5 | 19 |
| 7.18 | 18.9 | 18 |
| **7.45** | **20.7** | **15** |
| 7.80 | 37.1 | 5 |
| 8.30 | 22.2 | 14 |
| **8.46** | **42.4** | **4** |
| 8.87 | 19.6 | 16 |
| 9.21 | 24.8 | 12 |
| **9.44** | **22.6** | **13** |
| 9.95 | 18.2 | 20 |
| 10.93 | 36.4 | 7 |
| **11.35** | **65.5** | **1** |
| 11.91 | 19.1 | 17 |
| **12.35** | **31.7** | **9** |
| **12.63** | **61.4** | **2** |
| 12.86 | 32.2 | 8 |
| **13.67** | **31.1** | **10** |
| 13.90 | 25.4 | 11 |
| 14.14 | 37.1 | 6 |
| 15.00 | 51.3 | 3 |

## 4. Comparison of Super-Kamiokande and Gsi Power Spectra

We see from Figure 1 and Table 1 that the oscillation at 9.43 year$^{-1}$ is evident in *both* Super-Kamiokande and GSI power spectra. We can examine the significance of this correspondence by means of a combined analysis of both power spectra, using procedures designed for such a comparison [17]. We use the *Minimum Power Statistic* to search for frequencies for which *both* Super-Kamiokande and GSI show significant power. If each power is distributed exponentially, the following function is also distributed exponentially:

$$U(S_1, S_2) = 2 * Min(S_1, S_2) \quad , \tag{2}$$

where *Min(x,y)* indicates the smaller of *x* and *y*. This function, formed by combining the Super-Kamiokande power spectrum and the GSI-midnight power spectrum, is shown in Figure 4. Not surprisingly, the peak is found, with *U = 26.64*, at 9.43 year$^{-1}$.

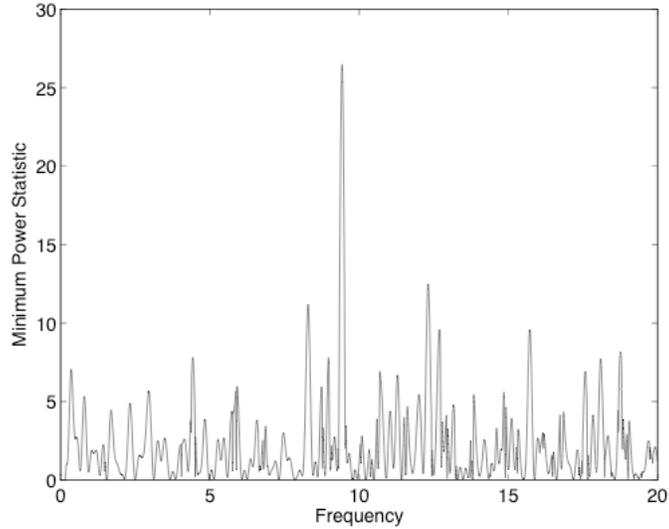

Figure 4. The minimum power statistic formed from the Super-Kamiokande power spectrum and the GSI midnight power spectrum. The most significant feature is found at 9.43 year$^{-1}$, with value 26.64, which could have occurred by chance in the band 0 – 20 year$^{-1}$ from normally distributed random noise in both datasets with a probability of 3.2 10$^{-10}$.

We may determine the corresponding probability from Equation 1. According to this equation, there is a probability of 2.7 10$^{-12}$ that this feature could have occurred by chance (from random noise in each of the datasets) at a specified frequency. Since there are 120 independent peaks in the band 0 – 20 year$^{-1}$, there is a probability of 3.2 10$^{-10}$ of finding by chance a value of $U$ of 26.64 or more anywhere in that band.

We have examined this correspondence in more detail by computing the amplitude and phase of each oscillation, with the results shown in Table 2. This relationship is shown schematically in Figure 5. For two waveforms that have completely different origins, we find a remarkable congruence. The agreement of the amplitudes and phases would seem to imply that the detected oscillation in the radon decay-product is due *entirely* to the detected oscillation in the neutrino flux.

Table 2. The frequency, amplitude and phase of the (nominally) 9.43 year$^{-1}$ oscillation, as it occurs in Super-Kamiokande measurements and GSI measurements.

|  | Super-Kamiokande neutrino measurements | GSI radon-decay measurements |
|---|---|---|
| Frequency | 9.43 ± 0.04 year$^{-1}$ (SK), | 9.44 ± 0.04 year$^{-1}$ (GSI); |
| Amplitude | 6.8 ± 1.7 % (SK), | 7.0 ± 1.0 % (GSI); |
| Phase | 124 ± 15 deg. (SK), | 124 ± 9 deg. (GSI). |

This result raises the possibility that – more generally - variations in the ambient neutrino flux may lead to variations in the beta decay process, conceivably that neutrinos are the principle – perhaps the only – cause of beta-decay variations. The influence may or may not be such as to change the decay rate; an influence that changes only the direction of travel of the emergent decay products, perhaps by a collective process, appears to be compatible with observational data.

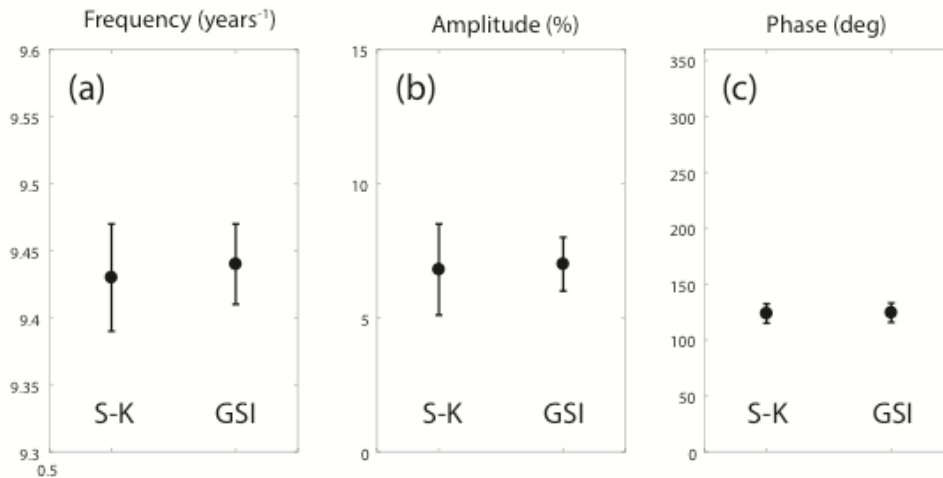

Figure 5. Frequency, amplitude and phase of the 9.43 year$^{-1}$ oscillation as it appears in Super-Kamiokande (SK) measurements and in GSI measurements.

It is notable that the oscillation has retained a constant phase from the beginning of the Super-Kamiokande measurements (in 1996) to the end of the GSI measurements (in 2017), a period of over 20 years. This phase stability is indicative of a "high-Q" oscillation, which is suggestive of the influence of a rotator – possibly the solar core.

**5. Hypotheses**
Although there have been proposals that some nuclear decay processes may be influenced by neutrinos [4 - 7], the close similarity of an oscillation detected in neutrino measurements and a corresponding oscillation detected in nuclear decay measurements has – to the best of our knowledge – not being anticipated.

There is obviously a need to reproduce (or attempt to reproduce) this result. If confirmed, it would establish that nuclear decays may indeed be influenced by neutrinos, leading to the challenge of determining the precise nature and mechanism of that influence.

It will hopefully be possible to obtain additional relevant information from measurements acquired at various neutrino observatories and from various nuclear-decay experiments. However, as a first step towards a theoretical investigation of the current problem, we may attempt to identify relevant hypotheses. We propose for consideration the following four hypotheses:

Hypothesis 1. The apparent oscillations are due simply to chance patterns in uncorrelated data.

The application of Equation 1 to oscillations identified in Figures 1, 2, and 3 and Table 1 indicates that this hypothesis is untenable.

Hypothesis 2. The oscillations identified in Figures 2 and 3 and Table 1 are due simply to meteorological influences [18].

A comparison of spectrograms formed from radon decay data and from meteorological data indicates that this hypothesis also is untenable. (See Figure 15 of ref. [7].) Furthermore, in view of the construction and location of the Super-Kamiokande Observatory, the parallel hypothesis that the oscillation in Super-Kamiokande measurements is due to meteorological influences is also untenable.

Hypothesis 3. Whether or not a radioactive nucleus decays in a certain time interval may be influenced by the ambient neutrino flux and this influence is independent of the design (in particular, of the geometry) of the experiment.

According to this hypothesis, whether or not a radionuclide exhibits variability would be insensitive to the design of the experiment. However, different experiments appear to give very different responses for the same nuclide. Note, as an example, an experiment by Bellotti et al. [19]. The "radon in air" version of the experiment displayed unmistakable evidence of a diurnal oscillation (which appears to be identical to a diurnal oscillation in GSI measurements - see Figure 4 of [7]), but the "radon in liquid" version showed no evidence of such an oscillation.

Hypothesis 4. The detection of gamma rays associated with the beta decay of a radioactive nucleus is sensitive not only to the magnitude of the ambient neutrino flux, but also to the relationship of the flow direction of the ambient neutrino flux to the geometry of the experiment. For instance, the direction of emission of the gamma rays may be related to the flow direction of the ambient neutrino flux.

According to this hypothesis, if the direction of emission of gamma rays is related to the direction of propagation of the ambient neutrino flux, and if the ambient neutrino flux is anisotropic, an experiment that is insensitive to the direction of emission would not detect variability, but an experiment that is sensitive to the direction of emission might exhibit variability.

According to currently available information (including the Bellotti experiment [19]), this hypothesis appears to be tenable. It is also worth noting that the design of the GSI experiment (which exhibits highly significant oscillations with amplitudes of several percent) is such that there is no obstacle between the generation of gamma rays and their detection.

We propose to study these (and possibly other) hypotheses in subsequent articles. We also plan to address in later articles the significance of the annual sidebands, and the significance of the low-frequency (annual, biennial, etc.) oscillations.

**Acknowledgments**
We thank the Super-Kamiokande Consortium for making available the data analyzed in Section 2.  For many stimulating and educational conversations in the course of this project, PAS thanks Roger Blandford, Daniel Freedman, Douglas Gough, Jeffrey Scargle, Robert Wagoner and Guenther Walther.